\documentclass[twocolumn,prb,10pt,aps,superscriptaddress,floatfix]{revtex4-2}
\usepackage[english]{babel}
\usepackage{graphicx}
\usepackage[pdftex]{epsfig}
\usepackage{ragged2e}
\usepackage{verbatim}
\usepackage{amsmath,amssymb}
\usepackage{times}
\usepackage[colorlinks,linkcolor=blue,citecolor=blue,urlcolor=blue]{hyperref}
\let\revappendix\appendix 
\usepackage[capitalise]{cleveref}
\usepackage{color}
\usepackage[table,xcdraw]{xcolor}
\usepackage{epstopdf}
\usepackage{physics}
\usepackage[export]{adjustbox}
\usepackage{feynmf}
\usepackage{dsfont}
\usepackage{bm}

\newcommand{\1}{\hspace*{-1pt}}

\newcommand{\bmpsi}{\bm{\Psi}}
\newcommand{\bmG}{\bm{G}}
\newcommand{\Ma}{\ell}
\newcommand{\Mb}{r}
\newcommand{\iwn}{i \omega_n}
\newcommand{\w}{\omega}
\newcommand{\G}{\mathcal{G}}
\newcommand{\Z}{\mathcal{Z}}

\providecommand{\Ga}[1]{\Gamma^\Lambda_{a\ #1}}
\providecommand{\Gb}[1]{\Gamma^\Lambda_{b\ #1}}
\providecommand{\Gc}[1]{\Gamma^\Lambda_{c\ #1}}

\begin{document}
\title{Frustrated Quantum Spins at finite Temperature: Pseudo-Majorana functional RG approach}
\author{Nils Niggemann}
\affiliation{Dahlem Center for Complex Quantum Systems and Institut f\"ur Theoretische Physik, Freie Universit\"{a}t Berlin, Arnimallee 14, 14195 Berlin, Germany}
\author{Bj\"orn Sbierski}
\affiliation{Department  of  Physics,  University  of  California,  Berkeley,  California  94720,  USA}
\author{Johannes Reuther}
\affiliation{Dahlem Center for Complex Quantum Systems and Institut f\"ur Theoretische Physik, Freie Universit\"{a}t Berlin, Arnimallee 14, 14195 Berlin, Germany}
\affiliation{Helmholtz-Zentrum f\"{u}r Materialien und Energie, Hahn-Meitner-Platz 1, 14109 Berlin, Germany}
\date{\today}

\begin{abstract}
The pseudofermion functional renormalization group (PFFRG) method has proven to be a powerful numerical approach to treat frustrated quantum spin systems. In its usual implementation, however, the complex fermionic representation of spin operators introduces unphysical Hilbert space sectors which render an application at finite temperatures inaccurate. In this work we formulate a general functional renormalization group approach based on Majorana fermions to overcome these difficulties. We, particularly, implement spin operators via an $SO(3)$ symmetric Majorana representation which does not introduce any unphysical states and, hence, remains applicable to quantum spin models at finite temperatures. We apply this scheme, dubbed pseudo Majorana functional renormalization group (PMFRG) method, to frustrated Heisenberg models on small spin clusters as well as square and triangular lattices. Computing the finite temperature behavior of spin correlations and thermodynamic quantities such as free energy and heat capacity, we find good agreement with exact diagonalization and the high-temperature series expansion down to moderate temperatures. We observe a significantly enhanced accuracy of the PMFRG compared to the PFFRG at finite temperatures. More generally, we conclude that the development of functional renormalization group approaches with Majorana fermions considerably extends the scope of applicability of such methods.

\end{abstract}
\maketitle

\section{Introduction}
Finding numerical solutions of quantum many-body problems is one of the core disciplines in modern condensed matter theory. In a wide range of physical settings the problem amounts to analyze ground-state and finite-temperature phases of a system of interacting spins on a lattice. Even though the corresponding microscopic models are often conceptually simple, such as two-body Heisenberg spin Hamiltonians, they may harbor a colorful range of physical phenomena including exotic types of long-range orders \cite{LMM}, quantum phase transitions \cite{sachdev,vojta03} or quantum spin liquids \cite{anderson73,balents10,Savary2017}. While quantum spin phases are traditionally described in terms of broken or unbroken symmetries, a more modern understanding also includes concepts such as long-range entanglement or topological order \cite{wen90} and reaches out to applications in the context of quantum information processing \cite{bennett00}.

Despite the shifts of focus which the field has gone through in the recent decades, the accurate numerical treatment of interacting quantum spin systems remains a highly challenging and longstanding problem. In fact, none of the currently available numerical methods is able to ultimately determine the eigenstates of a generic spin model. For example, quantum Monte Carlo methods \cite{sandvik91,sandvik99} which enjoy the invaluable advantage that numerical errors are, in principle, only of statistical nature, suffer from the infamous sign problem when applied to frustrated spin systems. Similarly, density matrix renormalization group, matrix product, and tensor network approaches \cite{white92,schollwoeck05,schollwoeck11,verstraete08,orus14} have made tremendous progress in recent years and are the undisputed method of choice for a variety of spin systems (particularly in one dimension). On the other hand, the scaling of the entanglement entropy poses a serious challenge for such techniques in higher dimensions.

An alternative approach is based on functional renormalization group (FRG) concepts \cite{Wetterich1993,Kopietz,polchinski84} which are, in principle, oblivious to the system's dimensionality. In its standard fermionic formulation this technique has first been applied in the context of electronic Hubbard-like models \cite{MetznerRMP2012,platt13,halbolth00} where it has become an established tool to describe competing types of long-range orders. In addition, a more recently developed variant of the FRG \cite{ReutherFRG} specifically targets quantum spin systems. The key conceptual step of this latter technique is to express the spin operators in terms of auxiliary fermions \cite{Abrikosov}, justifying the name pseudofermion functional renormalization group (PFFRG). Within the last decade the PFFRG has been successfully applied to a wide range of spin systems \cite{ReutherFRG,reuther11,reuther11_2,reuther11_3,reuther11_4,singh12,reuther12,goettel12,suttner14,Reuther2014,reuther14,iqbal15,rousochatzakis15,iqbal16,balz16,iqbal16_2,buessen16,hering17,Baez2017,iqbal17,buessen18,BuessenPRB2018,roscher18,Ruck2018,iqbal18,keles18,keles18_2,iqbal19,hering19,ghosh19,ghosh19_2,Buessen2019,niggemann19,revelli19,Roscher2019,chillal20,revelli20,kiese20,iida20,Kiese2020,Thoenniss2020} and has constantly been extended and generalized. Today, the PFFRG is, hence, remarkably flexible with a scope of applicability comprising two dimensional \cite{ReutherFRG,reuther11,reuther11_2,reuther11_3,reuther11_4,singh12,reuther12,goettel12,suttner14,Reuther2014,reuther14,iqbal15,rousochatzakis15,balz16,iqbal16_2,hering17,Baez2017,BuessenPRB2018,roscher18,Ruck2018,keles18,keles18_2,hering19,Buessen2019,niggemann19,Roscher2019,revelli20,kiese20,iida20,Thoenniss2020} and three dimensional \cite{iqbal16,buessen16,iqbal17,buessen18,iqbal18,iqbal19,ghosh19,ghosh19_2,niggemann19,revelli19,chillal20,Kiese2020} quantum spin systems on arbitrary lattices, including complex frustrated and longer-range coupled networks \cite{keles18,keles18_2} with general isotropic or anisotropic \cite{Buessen2019} two-body spin interactions. Further recent developments concern the generalization to arbitrary spin magnitudes $S$ \cite{Baez2017} or higher spin symmetry groups $SU(N)$ \cite{BuessenPRB2018,roscher18,kiese20} and, on a more technical level, the implementation of multi-loop schemes \cite{Ruck2018,Kiese2020,Thoenniss2020}.

Despite its success in accurately determining ground state spin correlations, the PFFRG comes along with a well-known obstacle. The aforementioned pseudofermionic description introduces an enlargement of the Hilbert space associated with states that do not correspond to states of the physical spin system. These unphysical states typically appear at energies above the ground state energy of the spin system. Thus, on the level of zero-temperature investigations, this problem has been argued to be rather mild and can be treated by shifting unphysical states to higher energies \cite{Baez2017}. In a recent investigation of this problem, on the other hand, the average spin magnitude within the PFFRG was found to differ from the theoretically expected result even for higher loop orders \cite{Thoenniss2020}. More importantly, the enlarged Hilbert space has so far prohibited an application to finite temperatures.

This work aims at resolving issues due to unphysical spin states by modifying the PFFRG on a very fundamental level. Instead of using a complex fermionic spin representation, we employ a certain, so-called $SO(3)$ Majorana fermion rewriting of spin operators \cite{MartinMajoranas,TsvelikMajorana} which does not generate unphysical states but only introduces redundant Hilbert space sectors. This property distinguishes it from other Majorana representations \cite{FuMajorana} and as such makes it attractive as a first candidate for a Majorana-based spin FRG. We, accordingly, dub our approach pseudo Majorana functional renormalization group (PMFRG) method. This modification opens up various directions of investigation: $(i)$ Most importantly, the PMFRG becomes applicable to finite temperatures which only requires small methodological adjustments presented below. $(ii)$ As a side product, we discuss how to calculate thermodynamic quantities such as the free energy, energy and heat capacity which have so far not been studied within the PFFRG. $(iii)$ To the best of our knowledge, a Majorana-implementation of the FRG has so far not been published. Our developments below are formulated in a general way such that they are applicable to arbitrary Majorana models also outside the realm of quantum magnetism. $(iv)$ Certain spin models, most prominently the Kitaev honeycomb model \cite{Kitaev2006}, are exactly solvable when expressed in terms of Majorana fermions. Although Kitaev's spin representation differs from the one employed here, the exact solution is also obtainable within the $SO(3)$ Majorana representation \cite{FuMajorana} used here. Even though not the focus of this work, one may thus expect that the PMFRG performs better for Kitaev-type spin models and perturbations thereof as compared to the PFFRG.

Apart from the methodological focus of this work, we also present various applications of the PMFRG to simple quantum spin models allowing us to assess its accuracy. As a first benchmark test we treat small clusters of up to six interacting spins where our results can be straightforwardly compared with exact diagonalization. An overall finding is that the thermodynamic behavior of the spin correlations from PMFRG are surprisingly accurate and reproduce the exact result significantly better than PFFRG. It should be emphasized that despite the finite Hilbert space of our spin clusters, their treatment within PMFRG is still highly non-trivial and poses the same challenges as for infinite lattice systems. Indeed, due to the incorporation of various mean-field limits, one can expect that the FRG unfolds its full strength only in infinite spin systems of two and higher dimensions. This motivates us to move on to frustrated Heisenberg models on 2D square and triangular lattices where we, likewise, find good agreement of thermodynamic properties with other approaches. A persistent technical issue, however, occurs in the low temperature limit where PMFRG detects spurious divergencies of spin correlations. We interpret this behavior as an artifact of the redundant Hilbert space sectors in our Majorana representation. While such subtleties remain to be further studied we expect that our developments lay the groundwork for various future directions of research and significantly enlarge the scope of applicability of FRG approaches.

The remainder of this work is organized as follows: After briefly reviewing the key concepts of the PFFRG in Sec.~\ref{sec:PFFRG}, we discuss in detail the properties of the $SO(3)$ Majorana representation in Sec.~\ref{sec:SO3Rep}. Thereafter, Sec.~\ref{sec:MFRG} formulates a general functional renormalization group approach for Majorana systems. The specific implementation for Heisenberg spin models in $SO(3)$ Majorana representation is discussed in Section~\ref{sec:vertexParametrization} with a particular focus on the parametrization of vertex functions, taking into account the system's symmetries. The resulting RG flow equations are presented in Sec.~\ref{sec:PmFRG} and the computation of various physical observables is detailed in Sec.~\ref{sec:Observables}. The following Secs.~\ref{sec:Clusters} and \ref{sec:2dSystems} discuss applications to small interacting spin clusters as well as to square and triangular lattice models. The paper ends with a conclusion in Sec.~\ref{sec:Conclusion}.


\section{Basic concepts of the PFFRG}\label{sec:PFFRG}
As a preparation for the following sections, we first briefly review basic concepts and properties of the PFFRG approach without being exhaustive on all methodological details. For a more detailed and self-contained description, we refer the interested reader to Refs. \cite{ReutherFRG,Baez2017,BuessenPRB2018,Buessen2019}.

The PFFRG is capable of treating general two-body spin Hamiltonians; in this work, however, only Heisenberg models of the form
\begin{equation}
H=\sum_{(i,j)} J_{ij}\sum_\alpha S_i^\alpha S_j^\alpha \label{eq:HeisenbergH}
\end{equation}
will be considered, where $(i,j)$ refers to all possible pairings of sites and $S^\alpha_i$ is the $\alpha$ component of a spin-$1/2$ operator at site $i$. We note in passing that recently developed FRG approaches \cite{KriegPRB2019,Goll2019} directly take Eq.~(\ref{eq:HeisenbergH}) as a starting point. In contrast, the PFFRG treats the interacting fermionic model that results from representing the spin-1/2 operators via (pseudo-) fermions $f_{i a}$ (with $a=\uparrow,\downarrow$) \cite{Abrikosov}:
\begin{equation}
S^\alpha_{i}=\frac{1}{2} \sum_{a,b} f_{i a}^{\dagger}\sigma^\alpha_{a b}f_{ib}\;.\label{eq:pf}
\end{equation}
Here and in the following, we set $\hbar = k_B = 1$.
However, this representation is a valid rewriting of the spin operators only in the local subspace with $\sum_a f_{i a}^\dagger f_{i a} = 1 $ while states with zero or double fermionic occupancy are unphysical. Since these spurious states carry zero spin, they may be considered as voids in the spin system, associated with an excitation energy on the order of the exchange coupling. As a consequence, ground state properties are believed to be largely unaffected by unphysical states, such that at $T=0$ the PFFRG may be faithfully implemented with the simpler condition $\sum_a \langle f_{i a}^\dagger f_{i a} \rangle= 1$. Other approaches aiming to enforce the occupancy constraint more rigorously introduce an energy penalty for unphysical states \cite{Baez2017} or a particular form of an imaginary chemical potential \cite{PopovFedotov}. In either case, the unphysical states remain an obstacle for an application of the PFFRG, especially at finite temperatures. This motivates us to implement the FRG with the Majorana representation discussed in Sec.~\ref{sec:SO3Rep} where no unphysical states occur. 

The key benefit of the representation in Eq.~(\ref{eq:pf}) is that the resulting model becomes amenable to fermionic many-body techniques such as the FRG which is formulated in terms of irreducible fermionic vertex functions (``essential parts of correlation functions''). The centerpiece of the method is given by a hierarchy of flow equations reminiscent of one-loop diagrammatic perturbation theory which describe the change of vertex functions when a Matsubara-frequency cutoff parameter $\Lambda$, introduced in the bare Green function $G^{0,\Lambda}(i\omega)=G^{0}(i\omega)\Theta(|\omega|-\Lambda)$, is varied. The basic idea is that at the starting point $\Lambda=\infty$, the bare propagator vanishes and all vertex functions are trivially known. For a numerical solution of the flow equations down to $\Lambda=0$ (the cutoff-free physical case), a truncation of the formally exact hierarchy of flow equations, usually at the level of the four-point vertex, is necessary.

The four-point vertex is directly related to the (momentum resolved) static spin susceptibility which represents the central outcome of the PFFRG approach. The onset of magnetic ordering is signaled by a divergence of the susceptibility along the RG flow (which in a finite system typically reduces to a finite peak or a kink). Accordingly, non-magnetic (and possibly quantum spin liquid) phases are characterized by an RG flow that remains smooth down to the lowest accessible $\Lambda$ scales.

Due to the lack of a small parameter in the purely interacting pseudo-fermion Hamiltonian, the truncation of the flow equations is an - a priori - uncontrolled procedure. It can be shown, however, that within the usual truncation on the level of the four-point vertex, both quantum fluctuations and classical ordering tendencies are correctly described in leading orders of $1/N$ and $1/S$, respectively \cite{Baez2017,BuessenPRB2018}. Here $N$ and $S$ describe the artificial enlargement of the spin's symmetry group {[}$SU(2)\rightarrow SU(N)${]}
and the spin length {[}$1/2\rightarrow S${]}, respectively. In two very recent works, certain contributions of the six-point vertex have been taken into account using a multiloop extension \cite{Thoenniss2020, Kiese2020} equivalent to a solution of the parquet self-consistency equations \cite{kugler18PRL,Kugler2018,kugler18NJP}. The quantitative robustness of the results with respect to increasing loop orders was interpreted as further evidence for the accuracy of the PFFRG.


\section{\texorpdfstring{$SO(3)$}{e} Majorana Representation}
\label{sec:SO3Rep}
In this section we discuss the $SO(3)$ Majorana representation \cite{MartinMajoranas,TsvelikMajorana} for spin-1/2 in detail. For each spin $S^\alpha_i$ at site $i$, three different flavors $\alpha \in \{x,y,z\}$ of Majorana fermions $\eta^{\alpha \dagger}_i = \eta^\alpha_i $ are introduced.  They fulfill the anticommutation relations $ \{ \eta^\alpha_i,\eta^\beta_j \} = \delta_{ij} \delta^{\alpha \beta}$ which imply $(\eta_i^\alpha)^2=1/2$. The formal Hilbert space dimension per Majorana is $\sqrt{2}$ as appropriate for half a (complex) fermion. The spin operators $S^\alpha_i = -\frac{i}{2} \sum_{\beta \gamma} \varepsilon^{\alpha \beta \gamma} \eta^\beta_i \eta^\gamma_i$, more explicitly written as
\begin{equation}
    S^x_i = -i \eta^y_i \eta^z_i \text{,} \qquad  S^y_i = -i \eta^z_i \eta^x_i \text{,} \qquad  S^z_i = -i \eta^x_i \eta^y_i \text{,} \label{eq:MajoranaRep}
\end{equation}
can be easily checked to fulfill the spin-1/2 algebra
\begin{equation}
S_i^\alpha S_i^\beta = \frac{1}{4} \delta^{\alpha \beta} + \frac{i}{2} \sum_{\alpha \beta \gamma } \varepsilon^{\alpha \beta \gamma} S_i^\gamma \text{.} \label{eq:Spin12Algebra}
\end{equation}
As an example, a Heisenberg coupling term from Hamiltonian \eqref{eq:HeisenbergH} is represented as
\begin{equation}
\sum_\alpha S^\alpha_i S^\alpha_j = -(\eta_{i}^{y}\eta_{i}^{z}\eta_{j}^{y}\eta_{j}^{z}+\eta_{i}^{x}\eta_{i}^{z}\eta_{j}^{x}\eta_{j}^{z}+\eta_{i}^{x}\eta_{i}^{y}\eta_{j}^{x}\eta_{j}^{y}).
\end{equation}

As usual for auxiliary particle representations, the $SO(3)$ Majorana representation comes with a gauge freedom. The local $\mathds{Z}_2$ gauge transformation $\eta^\alpha_i \rightarrow \varepsilon_i \eta^\alpha_i$ with $\varepsilon_i = \pm 1$ leaves spin operators invariant since each spin consists of a product of exactly two Majoranas with equal lattice index. This gauge freedom is also relevant to understand the structure of the Majorana Hilbert space. To see this, define the Majorana operator
\begin{equation}
    \tau_i = -2 i \eta^x_i \eta^y_i \eta^z_i \text{,} \label{eq:tauOperator}
\end{equation}
which anticommutes with any $\tau_j$ from a different site $j\neq i$  and fulfills

\begin{align}
\tau_i \eta^\alpha_j = \begin{cases}
               \eta^\alpha_i \tau_i &\mbox{if } i = j\\
               -\eta^\alpha_j \tau_i  &\mbox{if } i \neq j\\
           \end{cases} \text{.}
\end{align}
Consequently, $\tau_i$ commutes with all spin operators and thus with any spin Hamil\-tonian. To construct a set of mutually commuting operators one needs to pair $\tau_i$ with another conserved Majorana operator.

One choice \cite{ShnirmanMajorana} is to define an additional Majorana $\eta_i^0$ per site, so that the parity $p_i =2i\tau_i \eta_i^0$ with eigenvalues $\pm1$ is a constant of motion. These eigenvalues split the local Majorana Hilbert space of dimension four into two dynamically decoupled two-dimensional parts each of which are in one-to-one correspondence to the original local spin Hilbert space. To invoke $\eta_i^0$ in the Hamiltonian, parity projection schemes are required that eventually lead to one of two alternative four-Majorana spin representations \cite{FuMajorana}. However, as stated above, we will avoid this additional complication in the remainder of this work.

An alternative, non-local pairing scheme which does not introduce additional degrees of freedom requires an even number of sites $N$ \cite{BiswasPRB2011}.
Given an arbitrary but fixed pairing of sites $(i,j)$, we can define the $N/2$ parities $p_{(i,j)}=2i\tau_i \tau_j=\pm1$. Similar to above, each eigenstate of a spin Hamiltonian is $2^{N/2}$-fold degenerate, each copy labeled by the above parities. In other words, the total Majorana Hilbert space dimension of $2^{3N/2}$ is organized into the usual $2^N$ physical spin configurations, each with an artificial degeneracy of $2^{N/2}$. Choosing a different pairing of sites corresponds to a unitary rotation of the $2^{N/2}$ basis vectors for the artificial part of the Hilbert space. Note that since \cref{eq:MajoranaRep} fully reproduces the correct spin algebra without the need for an additional constraint, this Hilbert-space enlargement introduces no unphysical states, but only exact copies of the physical spin states \cite{FuMajorana}. This degeneracy is closely connected to the aforementioned local $\mathds{Z}_2$ gauge symmetry: As the transformation $\tau_i \rightarrow -\tau_i$ flips the parity $p_{(i,j)}$, it switches between degenerate states of different parities.

For thermodynamic properties, the above degeneracy leads to the relation
$\mathcal{Z}_{pm}=2^{N/2}\mathcal{Z}$ between the exact
partition functions defined in spin and $SO(3)$ pseudo-Majorana (pm) Hilbert space. Thus, we have for the physical free energy per site, $f=-T\log\left(\mathcal{Z}\right)/N$,
\begin{eqnarray}
f & = & f_{pm}+\frac{T}{2}\log\left(2\right) \label{eq:f_vs_fpm}
\end{eqnarray}
where the first term $f_{pm}\equiv-\frac{T}{N}\log\left(\mathcal{Z}_{pm}\right)$
will be computed via PMFRG and the second term accounts for the redundancy inherent in the $SO(3)$ Majorana representation.

Any expectation values for spin operators (or correlators) $\mathcal{O}_s$ are easily computed in the Majorana representation as well. This follows from the observation that the Majorana version of such an operator, $\mathcal{O}_{pm}$, is diagonal in the parity sector and the same is true for any physical density matrix $\rho_{pm}$, like for example the Boltzmann factor $\rho_{pm}\sim e^{-\beta H_{pm}}$. Then the degeneracy factor $2^{N/2}$ simply cancels \cite{SchadMajoranaNoRedundancy} and we have
\begin{equation}
\left\langle \mathcal{O}_{s}\right\rangle \equiv \frac{\mathrm{tr}\,\mathcal{O}_{s}\rho_{s}}{\mathrm{tr}\,\rho_{s}}=    \frac{\mathrm{tr}\,\mathcal{O}_{pm}\rho_{pm}}{\mathrm{tr}\,\rho_{pm}} \equiv \left\langle \mathcal{O}_{pm}\right\rangle.
\end{equation}

Finally, we discuss the role of rotations in spin space.
In order to employ the global $SO(3)$ symmetry of the Heisenberg Hamiltonian in \cref{eq:HeisenbergH} later on, we demonstrate here that the three Majoranas transform under $SO(3)$ rotations like the coordinates of a physical vector. Using $\tau_i$, the spin operators can be re-expressed as
\begin{equation}
    S^\alpha_i = \tau_i \eta^\alpha_i \text{.} \label{eq:SO(3)Majorana}
\end{equation}
We may now consider the general $SO(3)$ transformation $\eta^\alpha_i \rightarrow  \sum_\beta R_{\alpha \beta} \eta^\beta_i$ with $R_{\alpha \beta} \in SO(3)$ being a three dimensional rotation matrix.
As $\tau_i$ is invariant under this transformation \cite{FuMajorana}, spin operators must transform as
\begin{equation}
    R_{\alpha \beta} S^\beta_i = \tau_i \sum_\beta R_{\alpha \beta} \eta^\beta_i \text{.} \label{eq:SpinRotation}
\end{equation}
It follows that physical $SO(3)$ rotations of a spin $i$ are equivalent to rotations of the Majorana vector $\left(\eta^x_i,\eta^y_i,\eta^z_i\right)$. 

\section{General Majorana FRG Flow Equations}
\label{sec:MFRG}
As a basis for our FRG treatment of spin systems in pseudo-Majorana representation, we first introduce flow equations that are valid for general interacting Majorana Hamiltonians. To the best of our knowledge, such equations have not been published in the literature before. We consider
\begin{align}
 H = &\frac{i}{2} \sum_{\mu_{1,2}}{A_{\mu_1 \mu_2} \eta_{\mu_1}\eta_{\mu_2}} \nonumber\\
 + &\frac{1}{4!} \sum_{\mu_{1,2,3,4}}{V_{\mu_1 \mu_2 \mu_3 \mu_3} \eta_{\mu_1} \eta_{\mu_2} \eta_{\mu_3}\eta_{\mu_4}} \text{,} \label{eq:genMajoH}
\end{align}
where $\{ \mu_i \}$ is an arbitrary set of single-particle indices. As above, we use the convention $\left\{ \eta_{\mu_i} ,\eta_{\mu_j} \right\} = \delta_{\mu_i \mu_j}$. Majorana exchange statistics require the antisymmetry of $A$ and $V$ under exchange of any two indices, hermiticity mandates that both couplings must be real.

Assuming thermal equilibrium, we move on to an imaginary time path integral formulation \cite{SchadMajoranaNoRedundancy, NilssonPRB2013} defined in terms of Grassmann fields $\eta_\mu(\tau)$. The action reads
\begin{equation}
S = \int_0^\beta{
    d \tau \left( \sum_\mu{
        \frac{1}{2} \eta_\mu(\tau) \partial_\tau \eta_\mu(\tau) + H\left(\left\{ \eta_\mu(\tau) \right\}\right)}\right)} \text{,}
\end{equation}
where $\partial_\tau$ denotes a derivative with respect to imaginary time and $\beta=1/T$. We define the Fourier transform
$ \eta_\mu(\tau) = T \sum_{n}{e^{\iwn \tau} \eta_\mu(\iwn)} $
where the fermionic Matsubara frequencies are given by $\iwn = \pi T(2 n +1) $, with $n \in \mathbb{Z}$. In slight abuse of notation, in the following, we will denote $\omega_{n_1}$ by $\omega_1$ and equivalently for other frequencies. The non-interacting part of the action may then be written as
\begin{equation}
    S_0=- \frac{1}{2} \frac{1}{\beta^2} \sum_{\w_{1,2}}{ \sum_{\mu_{1,2}}{
        \eta_{\mu_1}(\w_1) \left[\bmG^{-1}_{0} \right]_{ \mu_1 \w_1,\ \mu_2 \w_2} \eta_{\mu_2}(\w_2)
        }}  \text{.} \label{eq:MajoranaActionGF}
\end{equation}
with the bare Majorana Green's function
\begin{align}
    \left[\bmG^{-1}_0\right]_{\mu_1 \w_1,\ \mu_2 \w_2} &= \left(i \w_1 \delta_{\mu_1 \mu_2} -i A_{\mu_1 \mu_2}\right) \beta\delta_{\w_1, - \w_2} 
    \text{.} \label{eq:Gzero}
\end{align}
This definition is analogous to the complex fermionic bare Green's function except for the opposing signs of the two frequencies in the Kronecker delta related to the absence of an independent Grassmann partner field $\bar{\eta}$ with a relative sign in the Fourier transform.

We are now ready to apply the general FRG scheme from Ref.~\cite{Kopietz}, derived for an action of a superfield vector $\bmpsi$ containing an arbitrary number of bosonic or Grassmann fields labeled by the composite index $l=(\w_l,\mu_l)$,
\begin{align}
    S[\bmpsi] &= S_0[\bmpsi] + S_{\text{int}}[\bmpsi] \nonumber\\
&= -\frac{1}{2} \int_l \int_{l'}{\Psi_l \left[\bmG^{-1}_0\right]_{l, l'}\Psi_{l'}}+ S_\text{int}[\bmpsi]\text{.} \label{eq:genAction}
\end{align}
where $\int_l=\beta^{-1}\sum_{\w_l}\sum_{\mu_l}$.
A comparison of \cref{eq:genAction} and \cref{eq:MajoranaActionGF} yields the direct correspondence $\bmpsi_{l=(\mu_l,\omega_l)}=\eta_{\mu_l}(\w_l)$. We emphasize the difference to the superfield vectors of complex fermions or bosons, which require an additional but independent superfield label, i.e.~$\bmpsi = (\bar{\psi}, \psi)$.

The starting point of the FRG scheme is the introduction of a cutoff scale $\Lambda$ in the bare Green's function $\bmG_0 \rightarrow \bmG_0^\Lambda$ such that $\bmG_0^{\Lambda=\infty} = 0$ and $\bmG_0^{\Lambda=0} = \bmG_0$. Although the flow equations describing the evolution of irreducible vertices with $\Lambda$ \cite{Kopietz} below are general, in the rest of this work, we will consider a multiplicative Matsubara frequency cutoff $\Theta^\Lambda(\w_1)$ to the bare Green's function
\begin{align}
   \left[\bmG^\Lambda_0\right]_{\mu_1 \w_1, \mu_2\w_2} &=\Theta^\Lambda(|\w_1|) \left[\bmG_0\right]_{\mu_1 \w_1, \mu_2\w_2}. \label{eq:Cutoff}
\end{align}
At zero temperature, this cutoff is often chosen to be a Heaviside function $\Theta^\Lambda(|\w|) = \theta(|\w| - \Lambda)$, at finite temperatures a smooth cutoff must be chosen instead. While a momentum based cutoff is also used in some works, we will not consider such schemes here, as our main focus lies on pseudo-Majoranas without kinetic energy.

As a consequence of the cutoff, the self-energy $\Sigma$ and the four-point vertex $\Gamma$ acquire implicit dependence on $\Lambda$. These quantities are defined via the Dyson equation in a superspace spanned by $(\w_i,\mu_i)$
\begin{equation}
    \bmG = \left[ \bmG_0^{-1}- \bm{\Sigma} \right]^{-1} \label{eq:Dyson}
\end{equation}
and the tree-expansion for the connected Green's functions
\begin{align}
    G^{4,c}_{l_1,l_2,l_3,l_4} = - \int_{l_{1',2',3',4'}}&{ \bmG_{l_1 l_{1'}} \bmG_{l_2 l_{2'}} \bmG_{l_3 l_{3'}} \bmG_{l_4 l_{4'}} }\nonumber\\
    &\times \Gamma_{l_{1'}l_{2'}l_{3'}l_{4'}}  \label{eq:TreeExpGamma}
\end{align}
respectively. This $\Lambda$-dependence is given by coupled differential equations, referred to as \textit{flow equations}. Physical results can be extracted from the solution at $\Lambda = 0$. Since the action for Majorana systems was rephrased in superfield notation, we can employ the associated general flow equations \cite{Kopietz} for $\Sigma^\Lambda$ and $\Gamma^\Lambda$. As appropriate in thermal equilibrium, and to simplify notation, we employ a modified version of the Green's function and vertices with the frequency conserving delta-function explicitly spelled out, 
\begin{subequations}
\begin{align}
    \bmG_{\mu_1 \w_1, \mu_2 \w_2} &= G_{\mu_1\mu_2}(\w_2) \beta \delta_{\w_1,-\w_2} \label{eq:Greenfunction}\\
    \bm{\Sigma}_{\mu_1 \w_1, \mu_2 \w_2} &= \Sigma_{\mu_1\mu_2}(\w_1) \beta \delta_{\w_1,-\w_2} \label{eq:Sigma}\\
    \Gamma_{\mu_1\w_1,\ \mu_2\w_2,\ \mu_3\w_3,\ \mu_4\w_4} &\equiv\Gamma_{\mu_1 \mu_2 \mu_3 \mu_4}(\w_1,\w_2,\w_3,\w_4 )\nonumber\\
    &\times  \beta \delta_{\w_1+\w_2+\w_3+\w_4,0} \text{.}
\end{align} 
\end{subequations}
With the above definition, the Dyson equation for fixed frequency indices, $\bmG_{-\w,\w} = \left[ \left[\bmG^{-1}_0\right]_{\w,-\w} - \bm{\Sigma}_{\w,-\w} \right]^{-1}$, can be written as 
\begin{align}
    G(\w) = \left[i \w -iA -\Sigma(\w)\right]^{-1} \text{.}
\end{align}
The Green's function and self-energy defined in Eq.~\eqref{eq:Greenfunction} and \eqref{eq:Sigma} fulfill $G(\w) = G^T(-\w)$ and $\Sigma(\w) = \Sigma^T(-\w)$, respectively.

We also restrict ourselves to the absence of parity symmetry breaking (expectation values of odd numbers of Majorana operators vanish) and neglect the contribution from the six-point vertex. The flow equation for the four-point vertex then separates into three distinct channels, each of which is characterized by one of the three bosonic transfer frequencies defined as
\begin{align}
s &= \w_1 + \w_2 = -\w_3 - \w_4, \nonumber\\
t &= \w_1 + \w_3 = -\w_2 - \w_4, \nonumber\\
u &= \w_1 + \w_4 = -\w_2 - \w_3 \text{.} \label{eq:MTransferFreq}
\end{align}
The Majorana flow equations for the interaction correction to the free energy, self energy and the four-point vertex read \cite{Kopietz}
\begin{widetext}
\begin{subequations}
\begin{align}
    \frac{d}{d\Lambda} F_{\text{int}}^{\Lambda} &= \frac{1}{2} \int_{\nu_{1,2,3,4}} T\sum_{\w'} S^{\Lambda}_{\nu_1 \nu_2}(\w')G^{0,\Lambda}_{\nu_2 \nu_3}(-\w') \left[G^{\Lambda}\right]^{-1}_{\nu_3 \nu_4}(-\w') \Sigma^\Lambda_{\nu_4,\nu_1}(\w')\label{eq:genMFlowEquationsGamma0}\\
    \frac{d}{d\Lambda} \Sigma^\Lambda_{\mu_1,\mu_2}(\w) &= -\frac{1}{2} \int_{\nu_{1,2}} T\sum_{\w'} S^\Lambda_{\nu_1 \nu_2}(\w') \Gamma^{\Lambda}_{\nu_1 \nu_2 \mu_1 \mu_2}(-\w',\w',\w,-\w) \label{eq:genMFlowEquationsSigma}\\
    \frac{d}{d\Lambda} \Gamma^\Lambda_{\mu_1,\mu_2,\mu_3,\mu_4}(\w_1,\w_2,\w_3,\w_4) &=
    \int_{\nu_{1,2,3,4}}T \sum_{\w} S^\Lambda_{\nu_1 \nu_2}(\w) \nonumber\\
    \times \bigg[& \Gamma^\Lambda_{ \mu_1 \mu_2 \nu_4 \nu_1}(\w_1, \w_2,\w -s ,-\w) \Gamma^\Lambda_{\nu_2 \nu_3 \mu_3 \mu_4}(\w ,-\w +s ,\w_3, \w_4)G^\Lambda_{\nu_3 \nu_4}(\w -s) \nonumber\\
    +&\Gamma^\Lambda_{\mu_1 \nu_1 \mu_3 \nu_4}(\w_1 ,-\w ,\w_3, \w-t) \Gamma^\Lambda_{\nu_2 \mu_2 \nu_3 \mu_4}(\w ,\w_2, -\w+t, \w_4) G^\Lambda_{\nu_3 \nu_4}(\w -t)\nonumber\\
    -&\Gamma^\Lambda_{\mu_1 \nu_4 \nu_1 \mu_4 }(\w_1 ,\w -u ,-\w, \w_4) \Gamma^\Lambda_{\nu_3 \mu_2 \mu_3 \nu_2}(-\w +u ,\w_2, \w_3, \w) G^\Lambda_{\nu_3 \nu_4}(\w -u) \bigg] \text{.} \label{eq:genMFlowEquationsGamma}
\end{align}
\label{eq:genMFlowEquations}
\end{subequations}
\end{widetext}
As the free energy does not feed back into the other flow equations it is usually not considered within FRG schemes. In this work, we use its solution to derive further thermodynamic quantities.
In these expressions, we have introduced the single-scale propagator which is defined as a matrix product of Green's functions
\begin{align}
\bm{S}^\Lambda &= - \bmG^\Lambda \left[ \frac{d}{d\Lambda}\left[ \bmG^\Lambda_0 \right]^{-1}  \right] \bmG^\Lambda \nonumber \\
S^\Lambda(\w_2) &= - G^\Lambda(\w_2) \left[ \frac{d}{d\Lambda}\left[ G^\Lambda_0 \right]^{-1}(\w_2)  \right] G^\Lambda(\w_2) \text{.}
\label{eq:SingleScale}
\end{align}
In order to solve the flow equations, initial conditions for self-energy and the four-point vertex are required. As the bare propagator vanishes in this limit, we immediately see that
\begin{align}
    F_{\text{int}}^{\Lambda \rightarrow \infty} &= 0, \nonumber\\
    \Sigma^{\Lambda \rightarrow \infty}_{\mu_1,\mu_2}(\w) &= 0, \nonumber\\
    \Gamma^{\Lambda \rightarrow \infty}_{\mu_1,\mu_2,\mu_3,\mu_4}(\w_1,\w_2,\w_3,\w_4) &= V_{\mu_1,\mu_2,\mu_3,\mu_4} \text{.}
\end{align}

\section{Symmetry-Based Vertex Parametrization}
\label{sec:vertexParametrization}

We now specialize the general Majorana FRG of this section to treat the interacting system of pseudo-Majoranas ensuing from the application of the representation \eqref{eq:MajoranaRep} to the Heisenberg spin-1/2 Hamiltonian \eqref{eq:HeisenbergH},
\begin{equation}
H = - \sum_{(i,j)} J_{ij} {\left(\eta_i^y\eta_j^y\eta_i^z\eta_j^z + \eta_i^z\eta_j^z\eta_i^x\eta_j^x+ \eta_i^x\eta_j^x\eta_i^y\eta_j^y \right)}. \label{eq:SO3Heisenberg}   
\end{equation}
As a first step, we proceed with a detailed discussion of the parametrization of vertices and propagators using the symmetries of our model. 
Following the approach of Ref.~\cite{Buessen2019}, we will first derive symmetry relations for the Green's functions defined as
\begin{align}
    G(1,2) &= \int_0^\beta{d \tau_1 d \tau_2 e^{i \w_1 \tau_1} e^{i \w_2 \tau_2} \expval{\eta_{\mu_1}(\tau_1) \eta_{\mu_2}(\tau_2) }} \nonumber\\
    &= G_{\mu_1,\mu_2}(\w_2) \beta \delta_{\w_1,-\w_2} \label{eq:G12} \\
    G^4(1,2,3,4) &= \int_0^\beta d \tau_1 d \tau_2 d \tau_3 d \tau_4 e^{i (\w_1 \tau_1+ \w_2 \tau_2 +\w_3 \tau_3+ \w_4 \tau_4)} \nonumber\\
    &\times\expval{\eta_{\mu_1}(\tau_1) \eta_{\mu_2}(\tau_2) \eta_{\mu_3}(\tau_3) \eta_{\mu_4}(\tau_4) } \label{eq:MGreensfct} \\
    &= G^4_{\mu_1,\mu_2,\mu_3,\mu_4}(s,t,u) \beta \delta_{\w_1+\w_2+\w_3+\w_4,0} \text{.} \label{eq:FreqParam}
\end{align}
where the labels $(1,2,3,4)$ contain all arguments that are not \textit{explicitly} specified, i.e $1 = (\mu_1,\w_1)$ in this case. Matsubara frequency conservation follows from the fact that thermal expectation values only depend on imaginary time differences. The time-ordering operator is suppressed since it is included in the path integral formalism by default. The properties derived in the following will then carry over to $\Sigma$ and $\Gamma$ due to their relations via \cref{eq:Dyson,eq:TreeExpGamma}.

\subsection{Hermiticity}
The Hamiltonian is a hermitian operator, satis\-fying $H = H^\dagger$. Due to $\expval{\mathcal{O}}^*  = \expval{\mathcal{O^\dagger}}$ and $\eta(\tau)^\dagger = e^{-H \tau} \eta e^{H \tau} = \eta(-\tau)$ in the Heisenberg picture, one can find the complex conjugate of the two-point Green's functions as
\begin{align}
G(1,2)^* &= \int{d \tau_1 d \tau_2 e^{- i \w_1 \tau_1 -i \w_2 \tau_2} \expval{ \eta_{\mu_2}(-\tau_2) \eta_{\mu_1}(-\tau_1)}}\nonumber\\
 &= -G(1,2) \text{.}
\end{align}
As a consequence, the two-point Green's function in Matsubara frequency space is purely imaginary and from an analogous argument, the four-point Green's function must be real
\begin{align}
&G(1,2) \in i \mathds{R}, \nonumber\\
&G^4(1,2,3,4) \in  \mathds{R} \text{.}
\end{align}

\subsection{Time reversal symmetry}
Time reversal $T$ is an anti-unitary operation $(\left\langle \psi | \psi^\prime \right\rangle^* = \left\langle T\psi | T\psi^\prime \right\rangle)$ which in the present case can be defined by performing a complex conjugation while leaving Majorana operators invariant \cite{Behrends2019}:
\begin{equation}
T i T^{-1} = -i \text{,} \:\:\quad
T \eta_\mu T^{-1} = \eta_\mu \text{.}
\end{equation}
This flips the sign of the spin operators \eqref{eq:MajoranaRep} as required. Time reversal symmetry is violated by an external magnetic field or, more generally, any Majorana bilinear in the Hamiltonian. For a $T$-symmetric Hamiltonian $THT^{-1}=H$, thermal expectation values obey $\left\langle \mathcal{O} \right\rangle = \left\langle T\mathcal{O}T^{-1} \right\rangle^*$. From this, we have $\expval{\eta_{\mu_1}(\tau_1) \eta_{\mu_2}(\tau_2)} = \expval{\eta_{\mu_1}(\tau_1) \eta_{\mu_2}(\tau_2)}^*$ and with Eq.~\eqref{eq:G12}, it follows that 
\begin{equation}
 G_{\mu_1 \mu_2}(\w_1, \w_2)= G_{\mu_1 \mu_2}(-\w_1, -\w_2)^* \text{.}
\end{equation}
Similarly, the four-point correlator has the property
\begin{equation}
 G^4(1,2,3,4) =  G^4_{\mu_1 \mu_2\mu_3 \mu_4}(-\w_1,-\w_2, -\w_3 , -\w_4)^* \text{.}
\end{equation}

\subsection{\texorpdfstring{Local $\mathds{Z}_2$}{e} gauge redundancy}
\begin{figure}
    \centering
    \includegraphics[width = \linewidth]{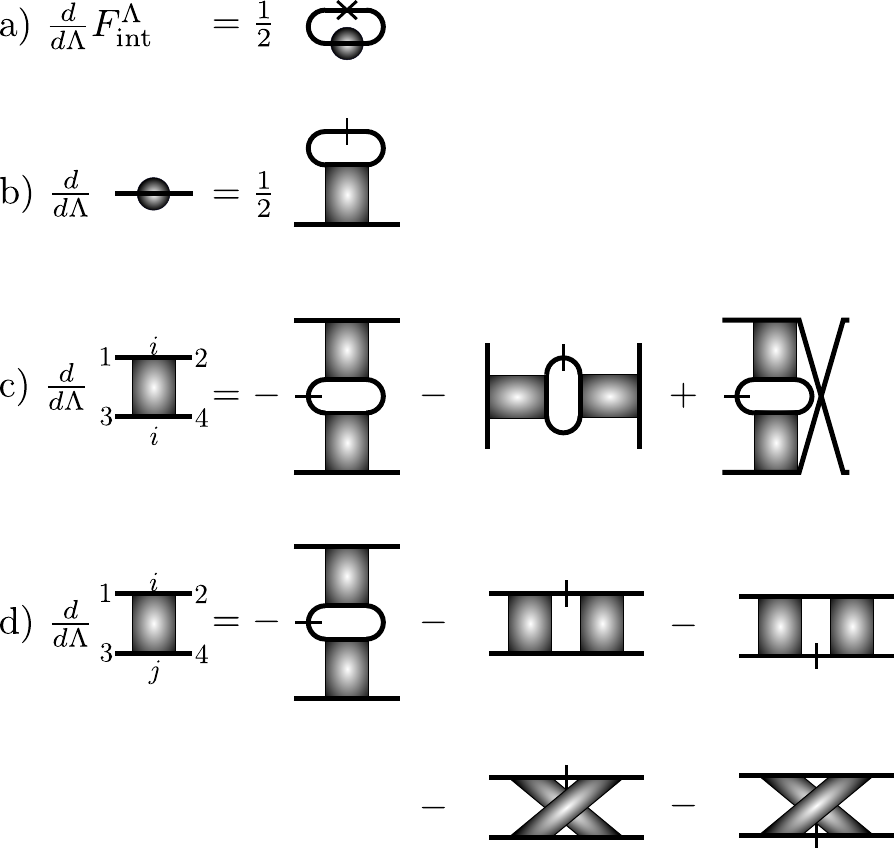}
    \caption{$\mathds{Z}_2$-invariant Majorana FRG flow equations for the interaction correction to the free energy (a), the self-energy (b), and the local (c) and nonlocal (d) four-point vertices.
    The order of labels $1=(\alpha_1,\w_1)$ always corresponds to that on the left hand side of the vertex flow equations such that the site index is conserved along solid lines. In these equations, internal lines correspond to fully dressed Green's functions $G_i(1,2)$, while the single scale propagator $S_i(1,2)$ is represented by a slashed line. Similarly, the crossed line in a) corresponds to the local propagator $\left[S G^{0} G^{-1} \right]_i(1,2)$.}
    \label{fig:Floweqs}
\end{figure}

Since our considerations from here on require the explicit specification of site indices, we will now separate the previously used superlabel $\mu$ into a site-index and a Majorana flavor $ \mu \rightarrow (i, \alpha)$.
In the $SO(3)$ Majorana representation spins are invariant under the gauge transformation $\eta^\alpha_i \rightarrow \varepsilon_i \eta^\alpha_i$ for all $\alpha=x,y,z$ with $\varepsilon_i = \pm 1$ for an arbitrary lattice site $i$. Since expectation values must be invariant under gauge transformations as well, we may write
\begin{equation}
    \expval{\eta^{ \alpha_1}_{i_1}(\tau_1) \eta^{ \alpha_2}_{i_2}(\tau_2)} = \varepsilon_{i_1}\varepsilon_{i_2} \expval{\eta^{ \alpha_1}_{i_1}(\tau_1) \eta^{ \alpha_2}_{i_2}(\tau_2)}\text{,}
\end{equation}
where $\varepsilon_{i_1}\varepsilon_{i_2} = -1$ may always be chosen for two different sites. As a consequence, non-zero propagators must contain an even number of Majorana operators from each site, so that
\begin{equation}
    G_{i_1 i_2}(1,2) \equiv  \delta_{i_1 i_2} G_{i_1}(1,2) \text{.}
\end{equation}
Likewise, the four-point correlator can only depend on up to two distinct sites only, so we choose
\begin{equation}
G^4_{i , i , j , j }(1,2,3,4) \equiv G^4_{ij}(1,2;3,4) \text{.} \label{eq:Msiteconvention}
\end{equation}
Correlators of the form $i j i j$ and $i j j i$ need to be brought to the standard form \cref{eq:Msiteconvention} using fermionic anticommutation rules, which restricts the number of allowed permutations in $G^4_{ij}(1,2;3,4)$ to exchanges of the first and last two indices only. As a consequence of the (bi-)local nature of propagators (four-point vertices), the site summations in the flow equations can be simplified. The special case $i =j$ for the four-point vertex needs to be considered separately. The corresponding flow equations can then be expressed diagrammatically as shown in \cref{fig:Floweqs}.
The bubble-diagram corresponding to the $s$-channel of the non-local vertex $\Gamma_{ij}$ shown in Fig. \ref{fig:Floweqs} d) is of particular interest. As in the PFFRG this diagram includes the random-phase approximation which is responsible for the emergence of long-range magnetic order \cite{Baez2017}.

\subsection{Lattice symmetries}
For simplicity, the systems that are considered in the following consist of equivalent sites. Correlators can then always be computed with one arbitrary reference site fixed. Combining this with local $\mathds{Z}_2$ gauge redundancy eliminates all site indices of the two-point correlator. Similarly, four-point correlators depend only on the distance vector between the two sites. Although this means that the order of site indices in $\Gamma_{ij}$ is irrelevant for systems with equivalent sites, we will not make use of this property. As a result, the pseudo-Majorana flow equations presented here are easily generalized towards non-Bravais lattices by adding an additional sublattice-index. Most lattice systems further exhibit point-group symmetries, such as the $C_4$ rotation symmetry and mirror planes of the square lattice, which can straightforwardly be used to reduce the numerical effort and are not further discussed in the following due to their lattice-specific nature.

\subsection{Global \texorpdfstring{$SO(3)$}{e} rotation symmetry}
\label{sec:SO3Symm}
The global $SO(3)$ spin-rotation symmetry of the Heisenberg model can easily be translated to vertex functions. As discussed in Sec.~\ref{sec:SO3Rep}, global spin rotations specified by a $3 \times 3$ rotation matrix $ R_{\alpha \mu}(\bm{\phi})$ act on the Majorana fermions as
\begin{equation}
    \eta^\alpha_i \rightarrow \sum_\beta R_{\alpha \beta}(\bm{\phi}) \eta^\beta_i \quad \forall i \text{.} \label{eq:MajoranaSO3Transf}
\end{equation}
The Heisenberg Hamiltonian is invariant under spin rotations due to the isotropic nature of its couplings.

We will now apply this symmetry to restrict the types of vertices and find relations between vertices with different flavor indices. Of particular interest are the specific rotations along the $x$, $y$ and $z$-axes as displayed in \cref{tab:symmetries}. The combination $R_x(\pi/2) \circ R_z(\pi/2) \equiv \mathcal{P}$ realizes an anti-cyclic permutation of the flavors.
\begin{table}
    \centering
\begin{tabular}{|c|c|c|c|}
\hline
 Angle & x & y & z \\
 \hline
 $\pi/2$ & $\eta^y \rightarrow -\eta^z$ & $\eta^x \rightarrow \eta^z$ & $\eta^x \rightarrow -\eta^y$ \\
 & $\eta^z \rightarrow \eta^y$ & $\eta^z \rightarrow -\eta^x$ & $\eta^y \rightarrow \eta^x$\\
 \hline
 $\pi$ & $\eta^y \rightarrow -\eta^y$ & $\eta^x \rightarrow -\eta^x$ & $\eta^x \rightarrow -\eta^x$ \\
 & $\eta^z \rightarrow -\eta^z$ & $\eta^z \rightarrow -\eta^z$ & $\eta^y \rightarrow -\eta^y$\\
 \hline
\end{tabular} \caption{Symmetry transformations corresponding to specific spin rotations along the $x,y$ and $z$ axes.} \label{tab:symmetries}
\end{table}
We apply these symmetries to correlators, using the convention $\gamma \neq \alpha \neq \beta \neq \gamma$ to refer to fixed, pairwise different flavors. In this way, we find that the two-point Green's function does not depend on any flavor labels.
\begin{align}
\expval{\eta^{\alpha}_1 \eta^{\beta}_2 } &\stackrel{R_\alpha(\pi)}{=} -\expval{\eta^{\alpha}_1 \eta^{\beta}_2 } = 0,\nonumber\\
\Rightarrow G_{\alpha_1,\alpha_2}(1,2) &= G_{\alpha_1}(1,2) \delta_{\alpha_1,\alpha_2} \stackrel{\mathcal{P}}{=}  G(1,2) \delta_{\alpha_1,\alpha_2} \text{.}
\end{align}
Because the four-point correlator has four flavor indices, at least two of them must be equal. An argument analogous to above shows that only vertices with an even number of flavors can be nonzero. Furthermore, rotations by $\pi/2$ transform different flavor combinations into each other, for instance
\begin{equation}
    \expval{\eta^{\alpha}_1 \eta^{\alpha}_2 \eta^{\beta}_3 \eta^{\beta}_4} \stackrel{R_\alpha(\pi/2)}{=}
    \expval{\eta^{\alpha}_1 \eta^{\alpha}_2 \eta^{\gamma}_3 \eta^{\gamma}_4} \text{.} \label{eq:G4labelswap}
\end{equation}
These arguments identify four independent flavor configurations for the four-point correlator, $G^4_{xxxx}(1,2,3,4)$, $G^4_{xxyy}(1,2,3,4)$, $G^4_{xyxy}(1,2,3,4)$ and $G^4_{xyyx}(1,2,3,4)$, all other types are either zero or related by \cref{eq:G4labelswap}.

After these simplifications, we consider a general rotation to find a relation between those four different correlators. Since they are now parametrized in terms of $x$ and $y$, we only need to consider rotations along the $z$-axis. The $\eta^x$ Majoranas then transform as $\eta^x_i \rightarrow \cos \theta \eta^x_i -\sin \theta \eta^y_i$ so that
\begin{equation}
    G^4_{xxxx} \stackrel{R_z(\theta)}{=} \expval{\left(\cos \theta \eta^x_1 -\sin \theta \eta^y_1 \right) \dots \left(\cos \theta \eta^x_4 -\sin \theta \eta^y_4 \right)} \text{.}
\end{equation}
Expanding the product and using the above symmetries, we obtain a relation independent of $\theta$
\begin{equation}
    G^4_{xxxx} = G^4_{xxyy} +G^4_{xyxy} +G^4_{xyyx} \text{,} \label{eq:SO3VerticesRelation}
\end{equation}
where the argument $(1,2,3,4)$ has been suppressed.
Since we considered an arbitrary rotation, our last consideration further serves as a proof that no other symmetries than the ones already shown may be found from $SO(3)$ rotations. Indeed, one arrives at the same identity regardless of which type of correlator one transforms (i.e.~transforming $G^4_{xyxy}$ yields the same result). Rotations along the $x$ or $y$ direction also generate no further information as a result of the permutation symmetry $\mathcal{P}$ and rotations around an arbitrary axis may always be decomposed as a product of $x,y$ and $z$ rotations.

\section{Pseudo-Majorana fRG Flow Equations}
\label{sec:PmFRG}
The symmetries of the last section imply the following parametrization of the pseudo-Majorana propagator,
\begin{equation}
G(1,2) = \G(-\w_1) \delta_{i_1, i_2} \delta_{\alpha_1, \alpha_2} \delta_{\w_1, -\w_2} \beta \text{,}
\end{equation}
where the imaginary and antisymmetric self-energy, abbreviated as $\Sigma(\w)=-i\gamma(\w)$, enters via the Dyson equation \eqref{eq:Dyson},
\begin{equation}
\G(\w)=  \frac{1}{i\w +\ i \gamma(\w)} \equiv - i g(\w)  \label{eq:MatsubaraGF}.
\end{equation}
In analogy to the real functions $\gamma(\w) $ and $g(\w)$ we also replace the imaginary single scale propagator via $S^\Lambda(\w)= -i \dot{g}^\Lambda(\w)$.
Due to the diagonal structure of the propagators, the symmetries for the four-point Green's functions then carry over to vertex functions [cf.~\cref{eq:TreeExpGamma}] whose frequency dependence is parametrized by the three bosonic frequencies introduced in \cref{eq:MTransferFreq}. The three independent four-point vertices are 
\begin{align}
  \Gamma_{a\ ij}(s,t,u) &\equiv \Gamma_{x i,\ x i,\ x j,\ x j}(s,t,u),\nonumber\\
  \Gamma_{b\ ij}(s,t,u) &\equiv \Gamma_{x i,\ x i,\ y j,\ y j}(s,t,u),\nonumber\\
  \Gamma_{c\ ij}(s,t,u) &\equiv \Gamma_{x i,\ y i,\ x j,\ y j}(s,t,u)\text{.} \label{eq:parametrization}
\end{align}
In the special case $i = j$, there are only two independent vertices since
\begin{equation}
\Gamma_{c\ i i}(s,t,u) = -\Gamma_{b\ i i}(t,s,u)\text{.} \label{eq:Gammacii}
\end{equation}
Vertices with negative bosonic frequencies are symmetry related to positive frequencies by time-reversal and a symmetry $t \leftrightarrow u$ further allows to reduce the numerical effort. Details are given in \cref{tab:stusymmetries}.  
\begin{table}[ht]
    \centering
\begin{tabular}{|c|c|c|}
\hline
 Operation & Symmetry for $\Gamma_{\mu\ ij}(s,t,u)$ &valid $\mu$\\
 \hline
 $1 \leftrightarrow 2 $ &  $t \leftrightarrow u \text{\ and\ }\Gamma_\mu \leftrightarrow -\Gamma_\mu$ & $a,b$ \\
 \hline
 $T \circ (1, 3) \leftrightarrow (2, 4) $  & $s \leftrightarrow -s$& $a,b,c$ \\
 $T \circ (1, 2) \leftrightarrow (3, 4)  $  & $t \leftrightarrow -t \text{\ and\ }i \leftrightarrow j$ & $a,b,c$ \\
 $T \circ (1, 2) \leftrightarrow (4, 3)  $ & $u \leftrightarrow -u \text{\ and\ }i \leftrightarrow j$ & $a,b,c$ \\
 \hline
\end{tabular} \caption{Transformations of the frequency arguments under time reversal $T$ and specific permutations of indices in$\Gamma_{ij}(1,2;3,4)$. The latter three rows apply to all three types of vertices and allow for a parametrization using positive frequencies only. Note that the final two permutations also exchange the order of $i$ and $j$ which is of importance for non-Bravais lattices. The remaining $t \leftrightarrow u$ symmetry for $\Gamma_c$ can be established by the exchange $1\leftrightarrow 2$, which changes the vertex to the form $\Gamma_{xyyx}$. Using \cref{eq:SO3VerticesRelation} to express $\Gamma_{xyyx}(s,t,u) = - \Gamma_c(s,u,t)$ in terms of the other vertices used in the parametrization, we obtain $\Gamma_{c\ ij}(s,u,t) =  (-\Gamma_{a\ ij}+\Gamma_{b\ ij}+\Gamma_{c\ ij})(s,t,u)$.}
\label{tab:stusymmetries}
\end{table}
In the above parametrization, the flow equations for the interaction correction to the free energy per spin and the self-energy may be simplified. Specifying the external flavor and site indices on the left hand side of the flow equations, we directly perform flavor sums to obtain
\begin{equation}
 \frac{d}{d \Lambda} f_{\text{int}}^{\Lambda} = - \frac{3T}{2}\sum_{\w} \dot{g}^\Lambda(\w) \frac{g^{0,\Lambda}(\w)}{g^\Lambda(\w)} \gamma^\Lambda(\w), \label{eq:MGamma0FlowFiniteT} 
\end{equation}
\begin{align}
  \frac{d}{d \Lambda} \gamma^\Lambda(\w_1) = \frac{T}{2}\sum_{\w} \sum_j \dot{g}^\Lambda(\w)  \bigg\{ 
   &\Gamma^{\Lambda}_{a\ i j}(0, \w_1+\w, \w_1- \w) \nonumber\\
  + 2 &\Gamma^{\Lambda}_{b\ i j}(0, \w_1+\w, \w_1- \w) \bigg\}\text{.} \label{eq:MSigmaFlowFiniteT}
\end{align}
Similarly, we may now express the flow equations for four-point vertices in the same way. For conciseness of notation, both the initial fermionic frequencies as well as the exchange frequencies $s,t$ and $u$ will be used on the right hand side which are defined by \cref{eq:MTransferFreq}, or inversely,
\begin{align}
\w_1 &= \frac{s+t+u}{2} \text{,} \quad \w_2 = \frac{s-t-u}{2} \nonumber\\
\w_3 &= \frac{-s+t-u}{2} \text{,} \quad \w_4 = \frac{-s-t+u}{2} \text{.}
\end{align}
\begin{widetext}
\begin{subequations}
\begin{align}
    \frac{d}{d\Lambda}\Gamma^\Lambda_{a\ i j}(s,t,u) &=  X^\Lambda_{a\ ij}(s,t,u)- \Tilde{X}^\Lambda_{a\ ij}(t,s,u)+ \Tilde{X}^\Lambda_{a\ ij}(u,s,t)\\
    \frac{d}{d\Lambda}\Gamma^\Lambda_{b\ i j}(s,t,u) &=  X^\Lambda_{b\ ij}(s,t,u)- \Tilde{X}^\Lambda_{c\ ij}(t,s,u)+ \Tilde{X}^\Lambda_{c\ ij}(u,s,t)\\
    \frac{d}{d\Lambda}\Gamma^\Lambda_{c\ i, j\neq i}(s,t,u) &=  X^\Lambda_{c\ ij}(s,t,u)- \Tilde{X}^\Lambda_{b\ ij}(t,s,u)+ \Tilde{X}^\Lambda_{d\ ij}(u,s,t)
\end{align} \label{eq:PMFRGFlowequations}
\end{subequations}
\begin{subequations}
    \begin{align}
        X^\Lambda_{a\ ij}(s,t,u) &= T \sum_\w \dot{g}^\Lambda(\w) g^\Lambda(\w+s) \sum_k \left[\Ga{ki}\left(s,\w +\w _1,\w +\w _2\right) \Ga{kj}\left(s,\w -\w _3,\w -\w _4\right)+ 2(a\rightarrow b)\right] \\
        X^\Lambda_{b\ ij}(s,t,u) &= T \sum_\w \dot{g}^\Lambda(\w) g^\Lambda(\w+s) \sum_k \left[\Ga{ki}\left(s,\w +\w _1,\w +\w _2\right) \Gb{kj}\left(s,\w -\w _3,\w -\w _4\right) + (a\rightarrow b)+ (a\leftrightarrow b)\right] \\
        X^\Lambda_{c\ ij}(s,t,u) &= T \sum_\w \dot{g}^\Lambda(\w) g^\Lambda(\w+s) \sum_k \left[\Gc{ki}\left(s,\w +\w _1,\w +\w _2\right) \Gc{kj}\left(s,\w -\w _3,\w -\w _4\right) + (\w_1 \leftrightarrow \w_2, \w_3 \leftrightarrow \w_4)\right] 
    \end{align} \label{eq:X}
    \end{subequations}
    \begin{subequations}
    \begin{align}
        \Tilde{X}^\Lambda_{a\ i, j\neq i}(s,t,u) = T \sum_\w \dot{g}^\Lambda(\w) g^\Lambda(\w+s) \big\{ \big[&\Ga{ij}\left(\w +\w _2,s,\w +\w _1\right) \Ga{ij}\left(\w -\w _4,s,\w -\w _3\right) \nonumber\\
        + &(\w_1 \leftrightarrow \w_2, \w_3 \leftrightarrow \w_4, i \leftrightarrow j)\big] +2(a\rightarrow c) \big\} \\ 
        \Tilde{X}^\Lambda_{b\ i, j\neq i}(s,t,u) = T \sum_\w \dot{g}^\Lambda(\w) g^\Lambda(\w+s) \big\{ \big[&\Ga{ij}\left(\w +\w _2,s,\w +\w _1\right) \Gc{ij}\left(\w -\w _4,s,\w -\w _3\right) \nonumber\\
        + &(\w_1 \leftrightarrow \w_2, \w_3 \leftrightarrow \w_4, i \leftrightarrow j)\big] + (a\rightarrow c)+ (a\leftrightarrow c)\big\} \\
        \Tilde{X}^\Lambda_{c\ i, j\neq i}(s,t,u) = T \sum_\w \dot{g}^\Lambda(\w) g^\Lambda(\w+s) \big\{ \big[&\Gb{ij}\left(\w +\w _2,\w +\w _1, s\right) \Gb{ij}\left(\w -\w _4,\w -\w _3, s\right) \nonumber\\
        + &(\w_1 \leftrightarrow \w_2, \w_3 \leftrightarrow \w_4, i \leftrightarrow j)\big] + (b\rightarrow c)\big\} \\
        \Tilde{X}^\Lambda_{d\ i, j\neq i}(s,t,u) = T \sum_\w \dot{g}^\Lambda(\w) g^\Lambda(\w+s) \big\{ \big[&\Gb{ij}\left(\w +\w _2,\w +\w _1,s\right) \Gc{ij}\left(\w -\w _4,\w -\w _3,s\right) \nonumber\\
        + &(\w_1 \leftrightarrow \w_2, \w_3 \leftrightarrow \w_4, i \leftrightarrow j)\big] + (b\leftrightarrow c)\big\} 
    \end{align} \label{eq:XTilde}
    \end{subequations}
\end{widetext}
To reduce the length of expressions, we have defined the single-channel contributions $X^\Lambda_{a,b,c\ ij}$ and $\Tilde{X}^\Lambda_{a,b,c,d\ ij}$ in \cref{eq:X,eq:XTilde} \cite{Ruck2018}.
The flow equations of local vertices are obtained noting that $\Tilde{X}^\Lambda_{a,b,c\ ii}(s,t,u) \equiv X^\Lambda_{a,b,c\ ii}(s,t,u)$.
We further stress that no flow equation for $\Gamma_{c\ ii}$ is required in \cref{eq:PMFRGFlowequations}, as this vertex is equivalent to $\Gamma_{b\ ii}$ by virtue of \cref{eq:Gammacii}.

In the PFFRG, the Katanin truncation scheme \cite{Katanin2004} was instrumental in providing sufficient feedback of the self-energy flow into the vertex flow equations \cite{ReutherFRG}. It amounts to promoting the single-scale propagator in the flow equations of four-point vertices to a full derivative of the Green's function
\begin{align}
S^\Lambda(\w) &\rightarrow \frac{d}{d \Lambda} \G^\Lambda(\w) \equiv S_{\text{conv.}}(\w) + S_\text{Kat}(\w) \nonumber \\
&= -\G(\w)^2 \frac{d}{d\Lambda} \left[\G^{0 \Lambda}(\w)\right]^{-1} + \G(\w)^2 \frac{d}{d\Lambda} \Sigma^\Lambda(\w) \text{.} \label{eq:SKatanin}
\end{align}
At zero temperature, frequencies become continuous and $T \sum_\w \rightarrow (2 \pi) ^{-1} \int d\w$. Using the sharp frequency cutoff $\G^{0 \Lambda}(\w) = \G^{0}(\w)\theta(|\w|-\Lambda)$, we thus obtain in the usual way using Morris's Lemma \cite{Morris1994}
\begin{equation}
  \dot{g}^\Lambda_{T=0}(\w)=  - \frac{\delta(|\w|-\Lambda)}{\w + \gamma^\Lambda(\w)} + \dot{g}^\Lambda_\text{Kat}(\w) \text{.} \label{eq:MSigmaFlowZeroT}
\end{equation}
At finite temperatures, a sharp cutoff of frequencies is no longer possible due to ambiguities that arise if $|\w| - \Lambda$ lies between two discrete Matsubara frequencies. Noting that there is still freedom in the choice of a smooth cutoff \cite{Roscher2019,Karrasch2008}, here we choose a Lorentzian cutoff function
\begin{equation}
\Theta^\Lambda(\w_n) = \frac{\w_n^2}{\w_n^2 + \Lambda^2} \text{.}
\end{equation}
Using \cref{eq:Dyson,eq:Cutoff,eq:SKatanin} the expressions for the Green's function and the single-scale propagator become
\begin{align}
g^\Lambda(\iwn) &= \frac{\w_n}{\w_n^2 + \w_n \gamma(\w_n) + \Lambda^2} \nonumber \\
\dot{g}^\Lambda(\iwn) &= -g^2(\iwn)\left(\frac{2 \Lambda}{\w_n} + \frac{d \gamma^\Lambda(\iwn)}{d\Lambda} \right) \text{.}
\end{align}
Finally, we need to specify the initial conditions for the newly defined vertices. After re-expressing the Heisenberg Hamiltonian \eqref{eq:HeisenbergH} by insertion of \cref{eq:MajoranaRep} for the spin operators, a comparison of coefficients yields 
\begin{align}
f_{\text{int}}^{\Lambda \rightarrow \infty} &= 0, \nonumber \\
\Sigma^{\Lambda\rightarrow \infty} &= 0, \nonumber\\
\Gamma_{a\ ij}^{\Lambda\rightarrow \infty}&=\Gamma_{b\ ij}^{\Lambda\rightarrow \infty} = 0, \nonumber\\
\Gamma_{c\ ij}^{\Lambda\rightarrow \infty} &= -J_{ij} \text{.}
\end{align}
To summarize, in our PMFRG scheme the flow equations for the free energy \eqref{eq:MGamma0FlowFiniteT}, self-energy \eqref{eq:MSigmaFlowFiniteT} and the vertex functions \eqref{eq:PMFRGFlowequations}, are solved numerically starting from large but finite $\Lambda\gg J$ down to $\Lambda \simeq 0$, approximating the initial conditions with the $\Lambda\rightarrow \infty$ values presented above. The flow of the free energy correction is integrated along the way but does not feed back into the other flow equations. Further details on the numerical implementation of the PMFRG are given in Appendix~\ref{sec:Numerics}. The next section describes how to extract observables along the flow and, most importantly, at the physical endpoint $\Lambda = 0$.

\section{Observables}
\label{sec:Observables}
In this section, we discuss the observables for Heisenberg spin-1/2 systems that will be studied in the following sections. These are the free energy, internal energy, heat capacity and static susceptibility. We explain how these observables are calculated from the eigenstates and -energies of the spin Hamiltonian \eqref{eq:HeisenbergH}, its exact representation with $SO(3)$ Majorana fermions and from the (approximate) solution of the PMFRG flow equations. 

From the partition function of a $N$-spin system with eigenenergies $E_n$, $\mathcal{Z}=\sum_{n}e^{-\beta E_{n}}$, the free energy per spin is given by
\begin{equation}
F/N=f=-\frac{T}{N}\log\left(\mathcal{Z}\right)=-\frac{T}{N}\log \sum_{n}e^{-\beta E_{n}}. \label{eq:f}
\end{equation}
The energy per spin is
\begin{equation}
E/N=-\frac{\partial\log\left(\mathcal{Z}\right)}{N\partial\beta}=\frac{\partial(f\beta)}{\partial\beta}=\frac{1}{N\mathcal{Z}}\sum_{n}E_{n}e^{-\beta E_{n}},    
\end{equation}
which as a function of $T$ also determines the heat capacity 
\begin{equation}
   C/N = \frac{\partial}{\partial T}E/N=\frac{1}{NT^{2}}\left(\frac{1}{Z}\sum_{n}E_{n}^{2}e^{-\beta E_{n}}-E^{2}\right).
\end{equation}
For small systems amenable to exact diagonalization, the rightmost expressions are most convenient. From the solution of the PMFRG flow equation \eqref{eq:MGamma0FlowFiniteT} for the interaction correction to the pseudo-Majorana free energy per site, we find $f_{pm}=f_{pm,0}+f_\text{int}^{\Lambda=0}$. The non-interacting free energy for three pseudo-Majoranas per site is $f_{pm,0}=-3T\, \mathrm{log}(2)/2$. Using the relation between $f_{pm}$ and $f$, Eq.~\eqref{eq:f_vs_fpm}, we finally obtain
\begin{equation}
f=-T\,\mathrm{log}(2)+f_\text{int}^{\Lambda=0}.
\end{equation}

The static spin-spin correlator can be computed from
\begin{equation}
    \chi_{ij} = \int_0^\beta d\tau \expval{S^z_i(\tau) S^z_j(0)}. \label{eq:MagSusc}
\end{equation}
Note that $\chi_{ij}$ can also be interpreted as a static (zero-field) susceptibility as it measures the response of a spin at site $i$ when a magnetic field is exerted at site $j$. We represent the spin operators by Majorana fermions and obtain from the vertices of the PMFRG at cutoff scale $\Lambda$,
\begin{align}
\chi^\Lambda_{ij} = &+ T^2 \sum_{\w_1 \w_2} g^\Lambda(\w_1)^2 g^\Lambda(\w_2)^2\Gamma^\Lambda_{c\ ij}(0,\w_1+\w_2,\w_1-\w_2) \nonumber\\
&+T\sum_{\w_1} g^\Lambda(\w_1)^2 \delta_{ij}\text{.}\label{eq:MFRGSusc}
\end{align}
Of particular interest for the two-dimensional systems below is the uniform susceptibility $\chi=\sum_{i,j}\chi_{ij}$.

\section{Application: Small Spin Clusters}
\label{sec:Clusters}

\begin{figure}
    \centering
    \includegraphics[width= \linewidth]{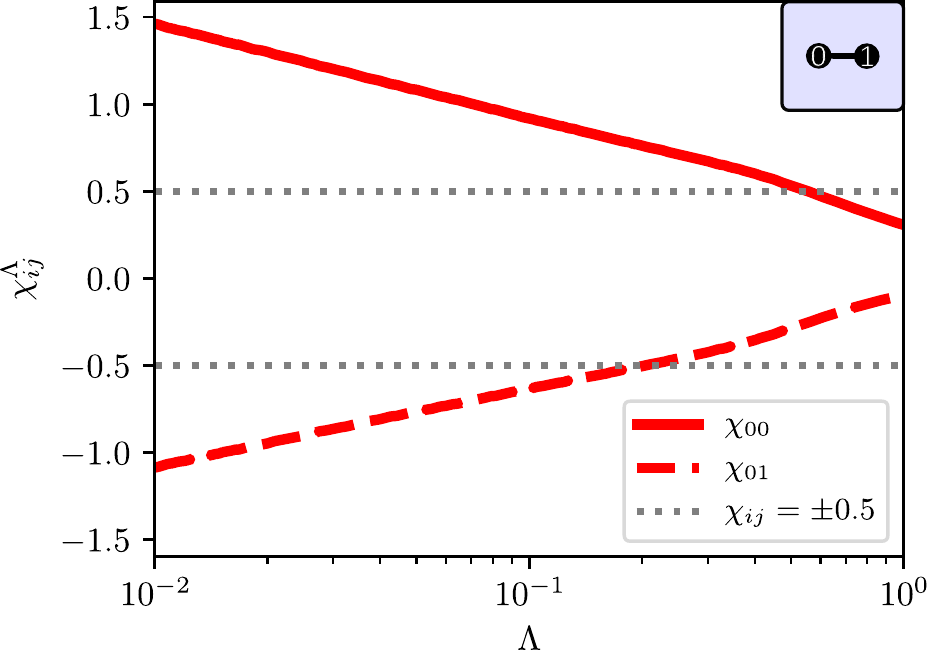}
    \caption{Zero temperature PMFRG flow of the static local and nonlocal susceptibilities $\chi_{ij}$ for the antiferromagnetic Heisenberg dimer. The grey dotted line represents the exact physical ($\Lambda=0$) result.
    }
    \label{fig:TzeroDimer}
\end{figure}

\subsection{Spin dimer and the fermion parity issue}
Small spin clusters constitute an ideal testbed for probing the accuracy of our approaches as they already represent non-trivial problems within the PMFRG (and PFFRG) but are still exactly solvable.
We first investigate the simple case of two spins, $i=0,1$ coupled with an antiferromagnetic Heisenberg interaction $J = 1$. 
Due to the small Hilbert space, this dimer model $H_{N=2}=\sum_\alpha S_0^\alpha S_1^\alpha$ is analytically solvable. While the free energy \cref{eq:f} is straightforwardly found, some care is required for the calculation of the susceptibility from the Lehmann representation where the term contributing in the case $i \nu +E_n -E_m = 0$ is often neglected in textbook derivations. We obtain
\begin{align}
\chi_{00} &= \frac{e^\beta -1 + \beta}{2(e^\beta +3)}, \nonumber\\
\chi_{01} &= - \frac{e^\beta  - 1 -\beta}{2(e^\beta +3)} \text{.} \label{eq:AnalyticalDimer}
\end{align}

Our PMFRG results for the static susceptibility  in the case $T=0$ are shown in \cref{fig:TzeroDimer} as a function of the cutoff. We find that $\chi_{ij}^\Lambda$ flows smoothly without any feature, surpasses the exact results $\chi_{ij} = \pm 0.5$ and diverges at $\Lambda = 0$. This unphysical divergence is not restricted to the Heisenberg dimer but appears in all other models considered here. However, the dimer allows for the most simple discussion of the origin of this divergence, which equally plagues the flow of the nonlocal vertices of type $\Gamma_{a,01}=\Gamma_{x0,x0,x1,x1}$ and $\Gamma_{c,01}= \Gamma_{x0,y0,x1,y1}$. 

To explain the origin of this divergence, consider the Heisenberg dimer which can be exactly solved in the $SO(3)$ Majorana representation,
\begin{equation}
H_{N=2} = -\frac{1}{4} p^x p^y p^z \left(p^x+ p^y+ p^z\right)\text{.} \label{eq:HDimer}
\end{equation}
Here, $p^\alpha \equiv 2 i \eta^\alpha_0 \eta^\alpha_1 $ are the three flavor parities related to the non-local parity introduced in Sec.~\ref{sec:SO3Rep} via $p_{(0,1)} = 2 i \tau_0 \tau_1 = -p^x p^y p^z$. While $p_{(i,j)}=\pm1$ is always conserved for generic spin systems, $p^\alpha=\pm1$ are additional constants of motion only for the dimer, \cref{eq:HDimer}. As any state, the ground state is $2^{N/2}=2$ fold degenerate and identified in this case by $p^\alpha=1$ or $p^\alpha=-1$ for all $\alpha$. Now consider the effect of a small perturbation, $H_{N=2}\rightarrow H_{N=2}+v p^x$. This does not correspond to any physical perturbation in terms of spin operators but lifts the ground state degeneracy. From this point of view, the ground state expectation value $\expval{p^\alpha}=0$ is fragile, any finite perturbation violating the conservation of $\tau_i$ as defined in \cref{eq:tauOperator} with $i=0,1$ generically causes $\expval{p^\alpha}=\pm1$. This effect is of course alleviated at finite temperature, where the relative population difference of the two lowest states split by $\sim v$ is controlled by the ratio $v/T$. Kubo's formula allows to formalize the above considerations for the linear response of $\expval{p^\alpha}$ with respect to $vp^x$,
\begin{equation}
\expval{p^\alpha} = - v G^R_{p^\alpha p^x}(i\w_k=0) \text{.} 
\end{equation}
In Matsubara frequency space, the retarded Green's function above may be obtained in the Lehmann representation noting that the parities are diagonal in the eigenbasis of the unperturbed Hamiltonian $\mel{n}{p^\alpha}{m} = p^\alpha_{n} \delta_{nm}$,
\begin{equation}
    \ G_{p^\alpha p^x}(i \w_k = 0) = \frac{\beta}{\Z} \sum_{n} e^{-\beta E_n} p^\alpha_{n}p^x_{n} \text{.}\label{eq:hoppingcorr}
\end{equation}
At low temperatures this yields $\beta = \frac{1}{T}$, similar to the Curie-like $1/T$ behaviour of the spin susceptibility of a free spin 1/2 which also features a degenerate ground state in the field-free case. 
In complete analogy to the spin susceptibility in Eq.~\eqref{eq:MFRGSusc}, we can now find the the tree expansion of the parity susceptibility $G_{p^\alpha p^x}(i \w_k = 0)$ in terms of the non-local vertices of type $\Gamma_a$ (for $\alpha=x$) or $\Gamma_c$ ($\alpha=y,z$). The expressions are similar to Eq.~\eqref{eq:MFRGSusc} but crucially probe different frequency combinations of the vertices ($t=0$ instead of $s=0$). In other words, non-local vertex components of order $\sim 1/T$ are inherently expected in the SO(3) Majorana representation. In an exact calculation, these components are responsible for the $1/T$ parity susceptibility of Eq.~\eqref{eq:hoppingcorr}, but do not affect the spin susceptiblity. However, the PMFRG is not an exact method and the unphysical behavior of $\chi^\Lambda_{ij}$ at $T=0$ must be a consequence of truncating the PMFRG flow equations which apparently causes this divergence to spill over to the spin susceptibility.
It is an interesting question if an improved two-loop truncation scheme (correct to order $\mathcal{O}(J^3))$ \cite{Ruck2018} or a recently developed but numerically demanding multi-loop generalizations of the (PF)FRG \cite{Kiese2020,Thoenniss2020}, can be a possible cure to this problem.

Fortunately, as the unphysical divergence in the PMFRG flow only occurs at $\Lambda=0$ and for $T=0$, there are other options to extract physically meaningful results without going beyond the flow equations presented above. First, it is still possible to detect magnetic phases, heralded by divergences at finite $\Lambda$ as we have tested for the $J_1-J_2$ square lattice Heisenberg model (data not shown).

\begin{figure}
    \centering
    \includegraphics[width= \linewidth]{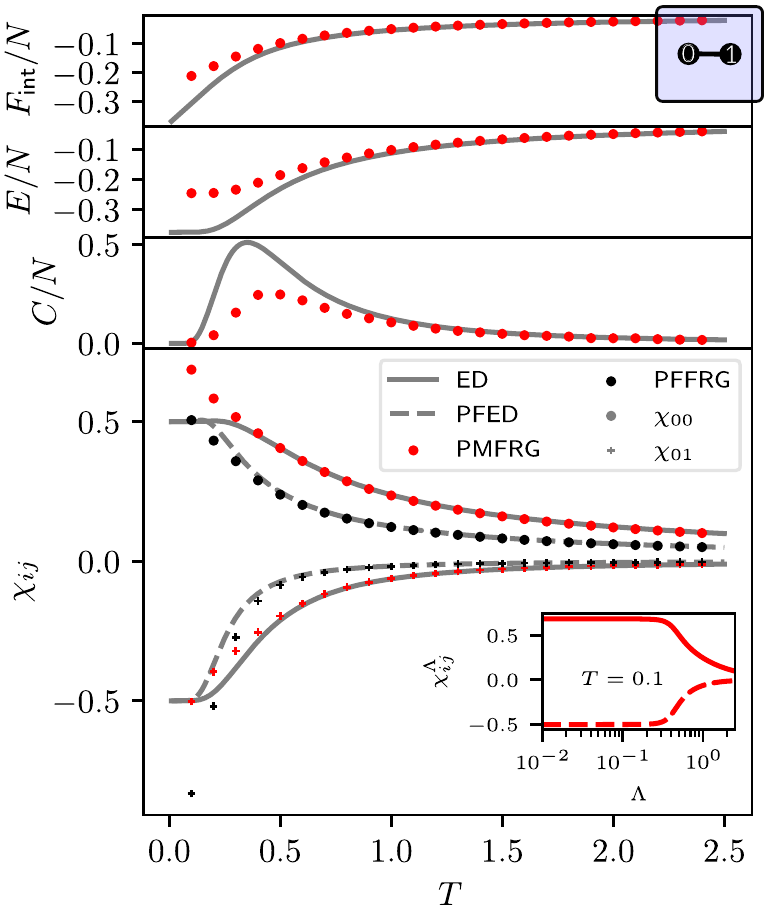}
    \caption{(Free) energy, heat capacity per spin and static susceptibilities of the Heisenberg dimer with $J=1$ obtained via PMFRG (red symbols) at $\Lambda=0$ as a function of temperature. Displayed in solid (dashed) grey lines are the results obtained by (pseudo-fermion) exact diagonalization, as well as the finite temperature spin susceptibilities of the PFFRG in black symbols. Each data point corresponds to a fully converged flow with respect to $\Lambda$ as demonstrated in the exemplary plot at $T=0.1$ (cf. \cref{fig:TzeroDimer}).
    }
    \label{fig:TSweep_Dimer}
\end{figure}

We devote the rest of the discussion to a second option, which is the restriction to finite temperatures. As explained above, this can be expected to suppresses the unphysical divergence and we indeed find all vertices and flowing susceptibilities converge towards $\Lambda \rightarrow 0$, see lower inset of \cref{fig:TSweep_Dimer} for $T=0.1$.

\subsection{Dimer and hexamer at finite temperature}
Results for the physical finite-$T$ susceptibility of the dimer at $\Lambda=0$ are shown in \cref{fig:TSweep_Dimer}. For $T \gtrsim 0.2$, we find a very close agreement between the susceptibility obtained via PMFRG and the exact result (solid lines) from \cref{eq:AnalyticalDimer}. The difference between the exact result and the PMFRG increases with decreasing temperature, in agreement with the discussion in the previous subsection.  We also show analogous results of the PFFRG, where the presence of unphysical states seriously compromises the accuracy of the results at any finite temperature scale. To support this interpretation, we have also included the results of an exact diagonalization scheme of the pseudo-fermionic Hamiltonian without projecting out unphysical states, further referred to as PFED. The close agreement between PFFRG and PFED demonstrates the problematic impact of unphysical states at finite temperatures which so far has no known resolution. One approach, the Popov-Fedotov projection scheme, suppresses unphysical states in exact calculations of observables upon the introduction of an imaginary chemical potential. However, producing a quarter-period shift of Matsubara frequencies \cite{PopovFedotov,Roscher2019}, this option has so far not been integrated in the PFFRG in a satisfactory manner.

Besides the magnetic susceptibility, our solution of the free energy flow equation enables us to compute a variety of related thermodynamic observables, such as the energy per spin and the heat capacity, also displayed in \cref{fig:TSweep_Dimer}. We observe good agreement at large enough temperatures. At intermediate scales $T \simeq 0.5$, the quality of the thermodynamic quantities from the PMFRG decreases as can be seen most clearly from the overestimation of the energy per spin or the underestimation of the peak in the heat capacity. These inaccuracies likely stem from the underestimation of the Majorana self-energy at small frequencies, a known problem in pseudo-fermion FRG approaches to spin systems of small dimensionality \cite{Reuther2014}.

\begin{figure}
    \centering
    \includegraphics[width= \linewidth]{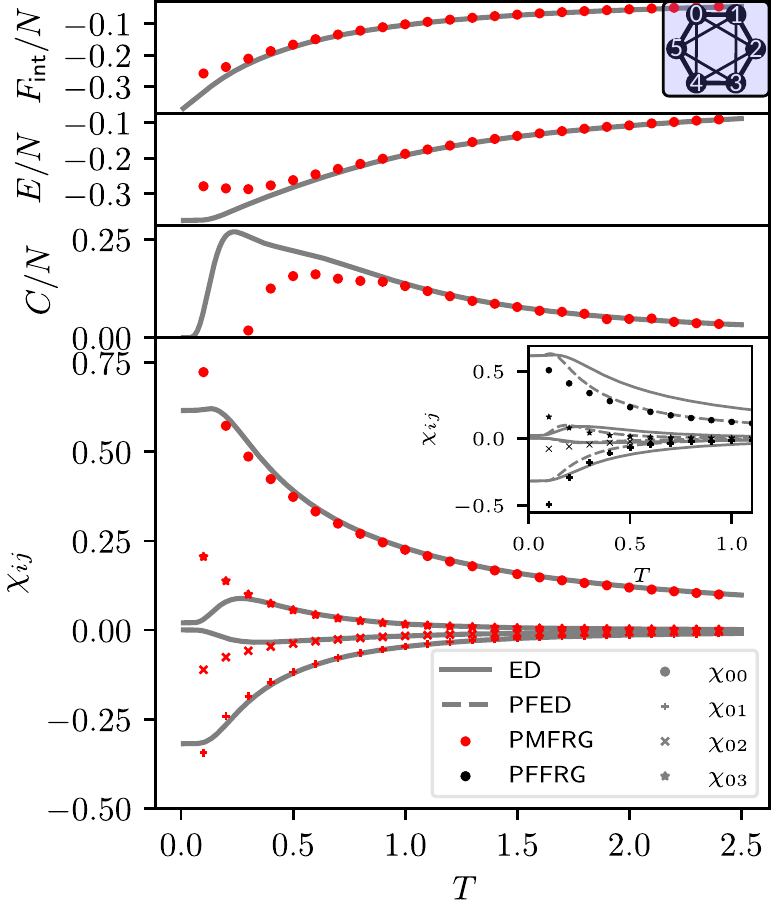}
    \caption{PMFRG results for the Heisenberg hexamer in analogy to \cref{fig:TSweep_Dimer}. The corresponding PFFRG and PFED results of the spin susceptibility are included in the inset.}
    \label{fig:TSweep_Hexamer}
\end{figure}

Analogous results are obtained for larger spin clusters such as the Heisenberg hexamer, a hexagon of six equivalent spins with nearest and next-nearest neighbor interactions, $J_1=1$ and $J_2 = 0.5$ respectively. As shown in \cref{fig:TSweep_Hexamer}, the PMFRG results are in good agreement with ED at not too small temperatures. The susceptibilities are generally more accurate than the thermodynamic properties. The susceptibility obtained via PFFRG shows large deviations from ED results at all temperatures.
We emphasize again that small spin clusters are particularly challenging within the FRG framework since its built-in mean-field limits are generally not expected to describe such systems accurately. On the other hand, mean-field approaches perform better in higher-dimensional systems. The FRG is, hence, expected to reach its full potential for larger or even infinite systems to which we move on in the following section.

\section{Application: Frustrated spin systems in 2D}
\label{sec:2dSystems}

We now turn to the application of the PMFRG to two-dimensional, frustrated and translational invariant Heisenberg spin models described by Hamiltonian \eqref{eq:HeisenbergH}. We first study the $J_1-J_2$ Heisenberg model on the square lattice with the parameter choice $J_2=0.5$ (where the system is expected to be non-magnetic) and then turn to the triangular lattice model with only nearest neighbor interaction, $J_2=0$. We work at finite temperature $T>0$ throughout and directly in the thermodynamic limit. Thus, as a technical modification from the previous section, we are required to limit the range of vertices to $|\mathbf{r}_i-\mathbf{r}_j| \leq L$, measured in units of the nearest-neighbor distance \cite{ReutherFRG}. Beyond this distance, vertices (and thus connected Green functions) are set to zero. We take $L \simeq 10$ large enough such that our results are converged in $L$. We study the same observables as in the previous section but report the uniform static susceptibility $\chi/N$ instead of $\chi_{ij}$. In contrast to the previous section, we plot these observables over $\beta=1/T$.

\begin{figure}
\centering
\includegraphics[width = \linewidth]{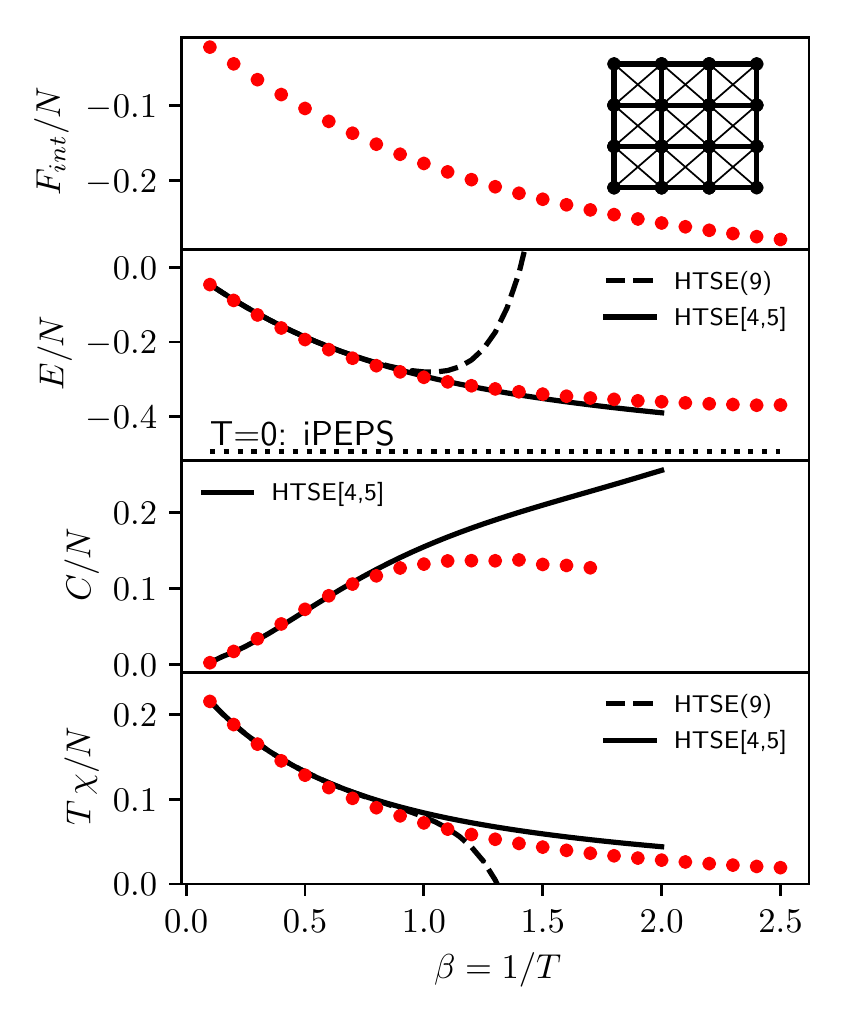}
\caption{PMFRG results (dots) for the $J_1-J_2$ square lattice Heisenberg model at $J_2/J_1=0.5$. The panels depict the single site contribution to the interaction correction to the free energy, internal energy, heat capacity and uniform susceptibility (top to bottom). The HTSE data (dashed line, up to 9th order) is taken from Ref. \cite{Rosner2003}, its 4,5 Pad\'e approximant is shown as a solid line. The iPEPS result for the ground state energy $E_0/N=-0.495$ from Ref. \cite{Poilblanc2017} is indicated as a dotted line.}
\label{fig:J1J2square}
\end{figure}

\begin{figure}
\centering
\includegraphics[width = \linewidth]{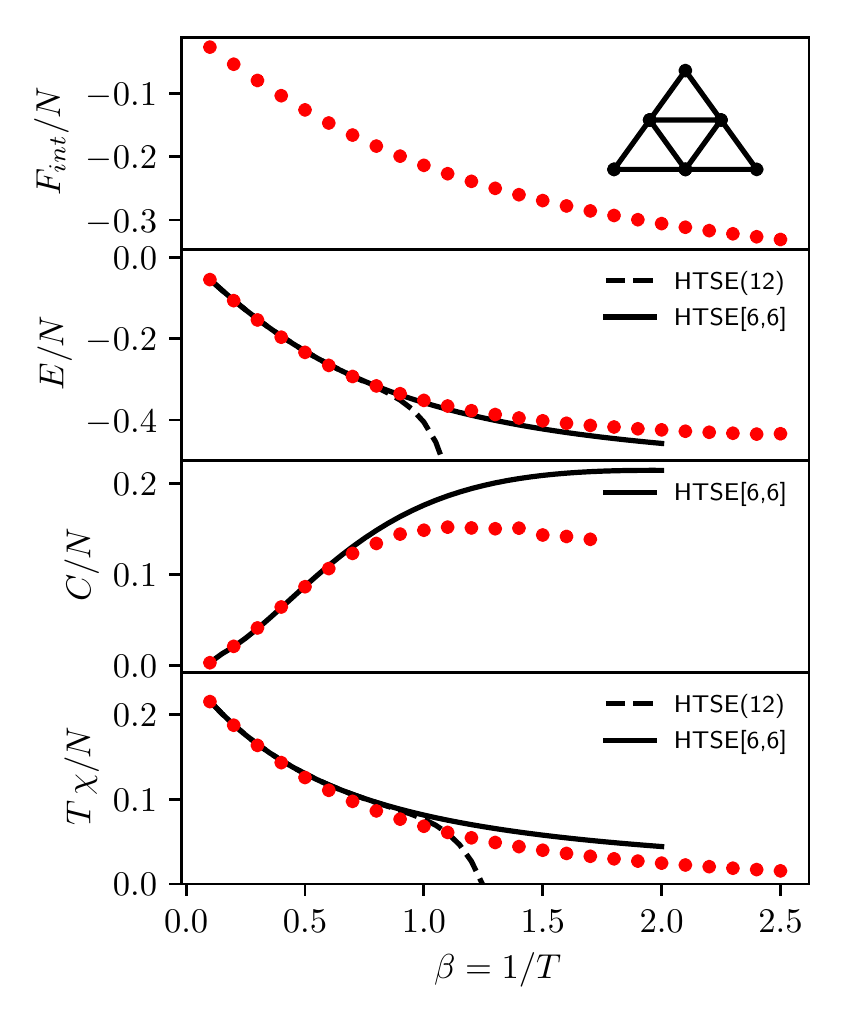}
\caption{PMFRG results (dots) for the nearest-neighbor triangular lattice Heisenberg model. The observables presented are analogous to Fig. \ref{fig:J1J2square}. The HTSE data (dashed line, up to 12th order) is taken from Ref. \cite{Elstner1993}, its 6,6 Pad\'e approximant is shown as a solid line.
}
\label{fig:triang}
\end{figure}

Our PMFRG results for the square lattice are shown in Fig. \ref{fig:J1J2square} (dots). We compare to the high-temperature series expansion (HTSE, dashed line) \cite{Rosner2003} and its 4,5 Pad\'e approximant (solid line) with an extended range of stability $\beta \lesssim 2$ \cite{Poilblanc2020}, to which our data is in reasonable agreement. We are not aware of $T>0$ tensor network results for the chosen model, but depict the iPEPS ground state energy $E_0/N=-0.495$ from Ref. \cite{Poilblanc2017} (dotted line). Finally, we remark that when applied to the unfrustrated nearest-neighbor Heisenberg model ($J_2=0$, data not shown), the PMFRG results agree only to the first order HTSE but deviate strongly from higher order and Monte Carlo data already for $T=1$. The likely reason is that for the current level of truncation of flow equations, the FRG is known to violate the Mermin-Wagner theorem \cite{Ruck2018}, and does, hence, not accurately capture the onset of magnetic order at $T=0$ in an unfrustrated Heisenberg system.

In Fig. \ref{fig:triang}, we show the PMFRG results for the triangular lattice nearest neighbor Heisenberg model (dots). Agreement to the HTSE data \cite{Elstner1993} (dashed line, up to 12th order) and its 6,6 Pad\'e approximant is similar as in the $J_1-J_2$ square lattice Heisenberg model of Fig. \ref{fig:J1J2square}. In the temperature range for which the Pad\'e-HTSE is shown, its accuracy was confirmed by recent experiments \cite{CuiPRB2018} and tensor network results \cite{ChenPRB2019}.

\section{Conclusion and outlook}
\label{sec:Conclusion}

In this work, we proposed a FRG approach to spin-1/2 quantum magnets with spin operators rewritten in the $SO(3)$ Majorana representation. Compared to the established PFFRG based on representing spins by complex fermions, our PMFRG method comes with a number of important conceptual differences, both on a technical level as well as regarding the scope for applications. First, as the Majorana nature of the spin representation is essential, we derived general FRG flow equations for generic interacting Majorana Hamiltonians. These  could potentially be useful for other applications \cite{Rahmani_2019}. Second, the $SO(3)$ Majorana representation avoids the unphysical states inherent in the complex fermion representation and instead features a redundant description of spin states reflected in a fixed artificial degeneracy. As a consequence, the truncation of flow equations is the only physical approximation made in the PMFRG. This explains why the PMFRG yields reasonably accurate results for finite temperatures, being out of reach for the PFFRG. In particular, we showed how the PMFRG can be used to compute thermodynamic quantities which are of great experimental relevance. On the downside, the PMFRG's precision at low temperatures suffers from a divergence of the $T=0$ flow, which we showed to be closely related to the (ground-)state degeneracy inherent in the $SO(3)$ Majorana representation, but ultimately caused due to inaccuracies introduced through the truncation of the hierarchy of flow equations. We thus conclude that, at the current stage, the PMFRG should be regarded not as a competitor to the PFFRG, but rather a complement in the practitioners toolbox tailored for finite and not too small temperatures.

Further work should investigate the potential of the recently proposed multiloop extension of the (PF)FRG \cite{Kugler2018,Thoenniss2020,Kiese2020} to mitigate the unphysical divergence mentioned above. Moreover, while the current paper has focused on Heisenberg systems with global spin rotation symmetry, generalization towards different classes of systems with reduced symmetries, i.e. Kitaev models and their variants, should be straightforward. Finally, we emphasize that the $SO(3)$ Majorana representation is only one out of several Majorana based spin representations \cite{FuMajorana}. Based on our results, we believe that these are promising but relatively underexplored venture points for the application of many-body methods in the study of spin systems. 

\section*{Acknowledgements}
We acknowledge useful discussions with Maxime Dupont, Christoph Karrasch, Alexander Penner, Achim Rosch, and Andreas Weichselbaum.
Computations were performed at the Curta cluster at Freie Universität Berlin and at the Lawrencium cluster at Lawrence Berkeley National Lab. BS acknowledges financial support by the German National Academy of Sciences Leopoldina through grant LPDS 2018-12. NN further acknowledges support from the German Research Foundation within the TR 183 (project A04).
%

\clearpage

\revappendix
    \section{Details on the numerical implementation}
    \label{sec:Numerics}
    The flow equations presented above can be solved using standard, error controlled Runge-Kutta schemes, such as the fifth-order Dormand-Prince method. In our case, we found little dependence of our results on the choice of the integration method used upon decreasing the relative and absolute accuracy to $\sim 10^{-2}$ or lower. In equivalence to implementations of the PFFRG, the maximum distance treated in four-point vertices $\Gamma_{ij}$ is limited to $|\mathbf{r}_i-\mathbf{r}_j| \leq L \simeq 10$ for translation invariant systems.
    
    At finite temperatures, we treat the frequency dependence by generating a set of $N_\w = 32$ positive Matsubara indices such that our results are converged in $N_\w$. The indices were chosen according to the following scheme such that the smallest frequencies are included exactly, while larger indices are more sparse and require for linear interpolation in between them:
    \begin{equation}
        n_i = \text{round}\left[z \sinh \left(\frac{i}{z}\right)\right] , \quad i = 0,1, \dots, N_\w \text{.} \label{eq:FTz}
    \end{equation}
    The parameter $z$ is then fully determined upon specification of the temperature, the number of frequencies, and the maximum frequency.
    Since the according (fermionic) frequencies are given by $\w_n = \pi T (2n +1)$, one needs to be careful when implementing fermionic symmetries such as $\gamma(-n_\w) = - \gamma(n_\w-1)$.
    Furthermore, the Matsubara integers corresponding to sums and differences of fermionic frequencies are
    \begin{align}
        \w_1 + \w_2 &\leftrightarrow n_{\w_1} + n_{\w_2} + 1 \nonumber\\
        \w_1 - \w_2 &\leftrightarrow n_{\w_1} - n_{\w_2}  \text{.}
    \end{align}
    As a result, only those sets of Matsubara integers that sum up to odd integers $n_s+n_t+n_u = 2n_{\w_1}+1$ are physical within energy conservation and will be evaluated in vertices.
    For the less robust implementation at $T=0$, we choose a logarithmic frequency mesh consisting of $N_\w = 96$ positive frequencies to avoid numerical errors from the finite frequency grid. The frequency integral in the Katanin contribution is then carried out numerically using a trapezoidal method.
\end{document}